\documentclass[twoside,12pt]{article}
\usepackage{indentfirst}
\usepackage{bm}
\usepackage{epsfig}

\begin{document}


\begin{center}
\LARGE\bf Fringe visibility and correlation in Mach-Zehnder interferometer with an asymmetric beam splitter
\end{center}

\footnotetext{\hspace*{-.45cm}\footnotesize $^*$This work was supported by Natural Science Foundation of China under
(Grant No.11975095); Shijiazhuang University
Doctoral Scientific Research Startup Fund Project (Grant No.20BS023).}
\footnotetext{\hspace*{-.45cm}\footnotesize $\ast$Corresponding author, E-mail:1102037@sjzc.edu.cn}

\begin{center}
\rm Yanjun Liu$^{\rm a}$, \ \ Meiya Wang$^{\rm a}$, \ \ Zhongcheng Xiang$^{\rm b}$, \ and \ Haibin Wu$^{\rm a,\ast}$
\end{center}

\begin{center}
\begin{footnotesize} \sl
${}^{\rm a}$ College of Physics, Mechanical and Electronical College, Shijiazhuang University, Shijiazhuang, 050035,
Hebei, China \\
${}^{\rm b}$ Institute of Physics, Chinese Academy of Sciences, Beijing 100190, China
\end{footnotesize}
\end{center}

\vspace*{2mm}

\begin{center}
\begin{minipage}{15.5cm}
\parindent 20pt\footnotesize
We study the wave-particle duality in a general Mach-Zehnder interferometer with an asymmetric beam splitter
from the viewpoint of quantum information theory.
The correlations(including the classical correlation and the quantum correlation) between the particle and the which-path detector are derived when they are
in pure state or mixed state at the output of Mach-Zehnder interferometer.
It is found that the fringe
visibility and the correlations
are effected by the asymmetric beam splitter and
the input state of the particle. The complementary relations between the fringe
visibility and the correlations are also presented.
\end{minipage}
\end{center}

\begin{center}
\begin{minipage}{15.5cm}
\begin{minipage}[t]{2.3cm}{\bf Keywords:}\end{minipage}
\begin{minipage}[t]{13.1cm}
wave-particle duality, fringe visibility, classical correlation, quantum correlation
\end{minipage}\par\vglue8pt

\end{minipage}
\end{center}

\section{Introduction}
A single quantum system has mutually exclusive properties, and these characteristics can be converted to each other
depending on the method of observation, which is known as Bohr's complementarity principle$[1]$. The well-known
example of the complementarity principle is wave-particle duality. A two-path interferometer, such as Young's double-slit
or Mach-Zehnder interferometer (MZI)$[2]$, is used to quantify the wave-particle duality. The wave-like property
and the particle-like property are shown by the fringe visibility and the which-path information (WPI) of the interferometer, respectively$[3-6]$.
If the path of the particle is known accurately, the fringe visibility will disappear. The more WPI is obtained,
the less the fringe visibility is. The complementarity
between the fringe visibility and the which-path knowledge has been studied greatly in theory and experiment$[7-30]$.¡¡

Originally, the uncertainty principle was attributed to the absence of interference fringe. Ever since the proposal that the entanglement
is responsible for the loss of interference$[31]$, a detector is introduced to the MZI, which is capable of making the path taken by the interfering
particle and called the which-path detector (WPD). The path information is obtained by reading the state of the detector. Although quantum entanglement
is the great interest to quantum mechanics and the great importance to quantum information processing, it cannot explain all non-classical correlation$[32-34]$, which is called quantum correlation (QC). QC are more general and more fundamental than entanglement. Thus, it is reasonable to believe that
not quantum entanglement but QC is responsible for the loss of interference. In this paper, we study both the wave-like property and
the particle-like property in a general MZI with an asymmetric beam splitter from the viewpoint of quantum information theory.
The fringe
visibility and the classical correlation (CC) or the QC
are used to quantify the wave-like behavior and the particle-like behavior, respectively.
The CC and the QC are calculated with either pure state or mixed state at the output of MZI, where by using the quantum discord (QD) represents the QC$[34]$. It is found that the fringe visibility and the correlations are effected by the asymmetric beam splitter(BS) and the
input state of the particle. The trade-off between the fringe visibility and the CC(QD) is also presented.

The paper is organized as follows. In Sec. \uppercase\expandafter{\romannumeral2}, we present the MZI with an asymmetric BS, and the state of the particle and the detector.
In Sec. \uppercase\expandafter{\romannumeral3}, we investigate the CC or QD of pure state and mixed state after the particle leaves
the asymmetry MZI. The complementary relationships of the fringe visibility and the correlations are also presented.
In Sec. \uppercase\expandafter{\romannumeral4}, we made our conclusion.

\section{\label{Sec:2}The setups and the state evolution}

\begin{figure} [htbp]
\centering
\includegraphics[clip=true,height=5cm,width=8cm]{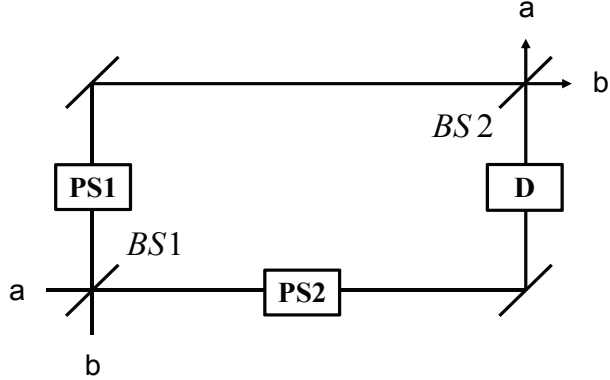}
\caption{The schematic sketch of the general Mach-Zehnder interferometer with the second BS asymmetric, a WPD is placed in path a.} \label{fig1.eps}
\end{figure}

A general MZI consists of two BSs and two phase shifters(PSs) as shown in Fig.~\ref{fig1.eps}. The incident particles are split into two
paths by the symmetrical beam splitter BS1. Orthogonal normalized states $|a\rangle$ and $|b\rangle$ are used to denote two possible paths, which support
a two-dimensional $H_{q}$. When the particles propagate in these two paths, PS1 and PS2 perform an rotation
\begin{equation}
U_{P}(\phi)=\exp ( -i\phi\sigma _{z})
\end{equation}
on the path qubit, where pauli matrix $\sigma_{z}=|b\rangle\langle b|-|a\rangle\langle a|$. Finally, these two paths are recombined by the asymmetric beam splitter BS2. The effect of the BS2 on the particle is denoted by
\begin{equation}
U_{B}(\beta)=\exp ( -i\frac{\beta }{2}\sigma _{y}),
\end{equation}
which is equivalent to performing a rotation around the $y$ axis by angle $\beta$. The BS2 is symmetrical when $\beta=\pi/2$.

To obtain the WPI, a WPD is placed on path a. When a particle initially in state
\begin{equation}
\rho _{in}^{Q}=\frac{1}{2}( 1+S_{x}\sigma _{x}+S_{y}\sigma
_{y}+S_{z}\sigma _{z}) \label{2eq-01}
\end{equation}%
go through the MZI, the operator $M=|b\rangle\langle b|\otimes I+|a\rangle\langle a|\otimes U$ is performed on the initial state
$\rho _{in}^{D}$ of the WPD, where $I$ and $U$ are the identical and unitary operator, respectively. In Eq.~(\ref{2eq-01}), the Bloch vector $\overrightarrow{S}=(\overrightarrow{S_{x}},\overrightarrow{S_{y}},\overrightarrow{S_{z}})$.
After the particle has passed through the MZI, the state becomes
\begin{eqnarray}
\rho _{f}^{QD}&=&U_{B}(\beta)MU_{P}(\varphi)U_{B}(\frac{\pi}{2})\rho_{in}^{Q}\rho_{in}^{D}
U_{B}^{\dagger}(\frac{\pi}{2})U_{P}^{\dagger}(\varphi)M^{\dagger}U_{B}^{\dagger}(\beta) \nonumber \\
&=&\frac{1}{4}( 1-S_{x}) ( 1+\sigma _{z}\cos \beta
+\sigma _{x}\sin \beta ) \otimes \rho _{in}^{D}  \nonumber \\
&&-\frac{1}{4}e^{-i\phi
}( S_{z}-iS_{y}) ( \sigma _{z}\sin
\beta -\sigma _{x}\cos \beta -i\sigma _{y}) \otimes \rho _{in}^{D}U^{\dagger} \nonumber \\
&&-\frac{1}{4}e^{i\phi}( S_{z}+iS_{y}) ( \sigma _{z}\sin \beta-\sigma _{x}\cos
\beta  +i\sigma _{y}) \otimes U\rho _{in}^{D} \nonumber \\
&&+\frac{1}{4}( 1+S_{x}) ( 1-\sigma _{z}\cos
\beta -\sigma _{x}\sin \beta ) \otimes
U\rho _{in}^{D}U^{\dagger},
\end{eqnarray}
and the reduced state of the WPD reads$[35]$
\begin{equation}
\rho^{D} _{f}= \omega_{b}\rho _{in}^{D}
+\omega_{a}U\rho _{in}^{D}U^{\dagger} \label{2eq-02},
\end{equation}
where
\begin{equation}
\omega _{a} =\frac{\cos ^{2}\left( \frac{\beta }{2}\right) \left(
1+S_{x}\right) }{1+S_{x}\cos \beta }, \ \ \ \ \ \  \omega _{b} =\frac{\sin ^{2}\left( \frac{\beta }{2}\right) \left(
1-S_{x}\right) }{1+S_{x}\cos \beta }.
\end{equation}
We note that the probabilities $\omega _{a}$ snd $\omega _{b}$ are only dependent on the parameters $S_{x}$ and $\beta$.

The probability that the particle is detected at output port $a$ reads
\begin{eqnarray}
p(\phi)&=&tr_{QD}[ \frac{1}{2}( 1-\sigma _{z}) \rho _{f}]  \nonumber \\
&=& \frac{1}{2}( 1+ S_{x}\cos\beta)
+\frac{A}{2}\sqrt{S_{z}^{2}+S_{y}^{2}}\sin\beta
\cos ( \alpha +\gamma +\phi )  \label{2eq-03},
\end{eqnarray}
where $A=|tr_{D}(U\rho _{in}^{D})|$, $\alpha$ and $\beta$ are the phases of $S_{x}+iS_{y}$ and $tr_{D}(U\rho_{in} ^{D})$, respectively. The
fringe visibility, which character the wave-like property of the particle, is defined via the probablity in Eq.~(\ref{2eq-03}) as
\begin{eqnarray}
V&\equiv&\frac{max P(\phi)-min P(\phi)}{max P(\phi)+min P(\phi)} \nonumber \\
&=&\frac{A\sin
\beta }{1+S_{x} \cos \beta }\sqrt{S_{z}^{2}+S_{y}^{2}},
\end{eqnarray}
where the maximum and minimum is achieved by adjusting $\phi$. We note that the expression of the
fringe visibility measured in either output port a or b is different in an asymmetric MZI,
however, the maximum of the fringe visibility remains the same no matter which port the particle is detected.
The fringe visibility obtains the maximum $A$
when the particle initially is in pure state and $\cos \beta=-S_{x}$$[35]$.

\section{\label{Sec:3} Information gain via general MZI}

To obtain the WPI, a measurement on the WPD is necessary. For the convenience of calculation, we assume that the initial state of the WPD is in pure state
denoted by $\rho _{in}^{D}= |r\rangle\langle r|$. For an arbitrary unitary operator U, states $| r\rangle$ and $| s\rangle \equiv U| r\rangle$
are linearly independent, which are either orthogonal state or non-orthogonal state. However, it is well known in quantum measurement theory that
non-orthogonal states cannot be accurately distinguished.
If conclusive results are made, errors are unavoidable. Thence it is desirable to distinguish these states with minimum probability of error, which is denoted by the
minimum error measurement$[36]$. Mathematically, the minimum error measurement for two-dimensional Hilbert space is characterized by the projective operators
$\Pi_{a}^{D}=|M_{a}\rangle\langle M_{a}|$ and $\Pi_{b}^{D}=|M_{b}\rangle\langle M_{b}|$. For state in Eq.~(\ref{2eq-02}), the basis of the minimum error
measurement read
\begin{small}
\begin{eqnarray}
\left\vert M_{a}\right\rangle  &=&-\frac{1}{A_{a}\sqrt{1-A^{2}}}(\frac{
1-\sqrt{1-4\omega _{a}\omega _{b}A^{2}}}{2\omega _{a}A}) \left\vert
r\right\rangle +\frac{1}{A_{a}\sqrt{1-A^{2}}}U\left\vert r\right\rangle  \nonumber \\
\left\vert M_{b}\right\rangle  &=&-\frac{1}{A_{b}\sqrt{1-A^{2}}}( \frac{
1+\sqrt{1-4\omega _{a}\omega _{b}A^{2}}}{2\omega _{a}A}) \left\vert
r\right\rangle +\frac{1}{A_{b}\sqrt{1-A^{2}}}U\left\vert \newline
r\right\rangle,   \label{3eq-01}           \nonumber \\
\end{eqnarray}
\end{small}
where
\begin{small}
\begin{eqnarray}
A_{a} &=&\sqrt{\frac{1-4\omega _{a}\omega _{b}A^{2}-\sqrt{1-4\omega
_{a}\omega _{b}A^{2}}( 1-2\omega _{a}A^{2}) }{2\omega
_{a}^{2}A^{2}( 1-A^{2}) }} \nonumber \\
A_{b} &=&\sqrt{\frac{1-4\omega _{a}\omega _{b}A^{2}+\sqrt{1-4\omega
_{a}\omega _{b}A^{2}}( 1-2\omega _{a}A^{2}) }{2\omega
_{a}^{2}A^{2}( 1-A^{2}) }}, \nonumber \\
\end{eqnarray}
\end{small}

In fact, the basis in Eq.~(\ref{3eq-01}) are the eigenstates of the operator $
\omega_{a}U\rho _{in}^{D}U^{\dagger}-\omega_{b}\rho _{in}^{D}$. In this section,
after the particle leaves the asymmetry MZI, we will study the CC and QD
between the particle and the WPD when they are in pure state or mixed state.

\subsection { Classical correlation via general MZI}

In quantum information theory, the CC$[37]$ between the particle and the WPD can be expressed as
\begin{equation}
\mathcal{J}(\rho^{QD})= max[S(\rho^{Q})-S(\rho^{Q}|\{\Pi_{k}^{D}\})]  \label{3eq-02},
\end{equation}
where the $\rho^{Q}$ is the reduced state of the particle, $S(\rho^{Q})$ is the von Neumann entropy, $S(\rho^{Q}|\{\Pi_{k}^{D}\}$ is the
quantum conditional entropy, $\Pi_{k}^{D}$ is the measurement operator performed on the WPD.

(1) Classical correlation for the pure state at the output of MZI

We assume that the particle is entangled with the WPD after the particle leaves the MZI. They are in pure state

\begin{equation}
\rho^{QD}_{1}=|\psi\rangle\langle\psi| \label{3eq-03},
\end{equation}
where $|\psi\rangle =\sqrt{ \omega_{a}}|a\rangle\otimes U|r\rangle+ \sqrt{ \omega_{b}}|b\rangle \otimes |r\rangle$.
Since $\rho^{QD}_{1}$ is in pure state, its quantum conditional entropy equal to zero. It also means that when $\rho^{QD}$ is pure state, the classical correlation between the particle and the WPD does not depend on the measurement
operator $\{\Pi_{k}^{D}\}$. The CC between the particle and the WPD reads
\begin{small}
\begin{eqnarray}
\mathcal{J}(\rho^{QD}_{1})&=&-\frac{B+\sqrt{B^{2}-\sin ^{2}\beta ( 1-S_{x}^{2}) (
1-A^{2}) }}{2B}
\log \frac{B+\sqrt{B^{2}-\sin ^{2}\beta (
1-S_{x}^{2}) ( 1-A^{2}) }}{2B}  \nonumber \\
&&-\frac{B-\sqrt{B^{2}-\sin ^{2}\beta ( 1-S_{x}^{2}) (
1-A^{2}) }}{2B}
\log \frac{B-\sqrt{B^{2}-\sin ^{2}\beta (
1-S_{x}^{2}) ( 1-A^{2}) }}{2B}, \label{3eq-04}
\end{eqnarray}
\end{small}
where $B=1+S_{x}\cos \beta$.

(2) Classical correlation for the mixed state at the output of MZI

The state of the WPD and the particle is always in a superposition
state due to the entanglement between them. A observer must subjectively select the state of the WPD and then read it
out to obtain the path information. In order to avoid the subjective selection, Zurek proposed "Environment-induced superselection rules"$[38]$,
which is equivalent to introducing the environment into the original system. We can obtain the state of the particle and the WPD
\begin{equation}
\rho_{2}^{QD} = \omega_{a}|a\rangle \langle a| \otimes U|r\rangle \langle r|U^{\dagger}+  \omega_{b}|b\rangle\langle b| \otimes |r\rangle\langle r|\label{3eq-05}
\end{equation}
by tracing over the degree of the environment. The state $|a\rangle(|b\rangle)$ is correlated with the state $U|r\rangle(|r\rangle)$ due to the "Environment-induced superselection". The observer can accurately know the state of the particle by reading out the state of the WPD.
If the WPD is in the state $U|r\rangle(|r\rangle)$, the state of the particle is $|a\rangle(|b\rangle)$.
Different from the $\mathcal{J}(\rho^{QD}_{1})$, the $\mathcal{J}(\rho^{QD}_{2})$
depends on the measurement basis. According to the numerical analysis,
when the measurement basis $\Pi_{k}^{D}=|M_{k}\rangle\langle M_{k}|$ induced by the environment is the same as the measurement
basis vector used in error-minimum measurement, the quantum condition entropy $S(\rho^{Q}|\{\Pi_{k}^{D}\})$ obtains the minimum.
The CC between the particle and the WPD reads
\begin{small}
\begin{eqnarray}
J( \rho _{2}^{QD})&=&-\omega _{b}\log \omega _{b} -\omega _{a}\log \omega
_{a} +\omega _{a}( S\sin \gamma -A\cos \gamma ) ^{2}
\log [
\frac{\omega _{a}( S\sin \gamma -A\cos \gamma ) ^{2}}{\omega
_{a}( S\sin \gamma -A\cos \gamma ) ^{2}+\omega _{b}\cos
^{2}\gamma }]  \nonumber \\
&&+\omega _{b}\cos ^{2}\gamma \log [ \frac{\omega _{b}\cos ^{2}\gamma }{%
\omega _{a}( S\sin \gamma -A\cos \gamma ) ^{2} +\omega _{b}\cos
^{2}\gamma }]
+\omega _{a}( A\sin \gamma +S\cos \gamma ) ^{2} \nonumber \\
&&\log [
\frac{\omega _{a}( A\sin \gamma +S\cos \gamma ) ^{2}}{\omega
_{a}( A\sin \gamma +S\cos \gamma ) ^{2}+\omega _{b}\sin
^{2}\gamma }]
+\omega _{b}\sin ^{2}\gamma \log [ \frac{\omega _{b}\sin ^{2}\gamma }{%
\omega _{a}( A\sin \gamma +S\cos \gamma ) ^{2}+\omega _{b}\sin
^{2}\gamma }],  \label{3eq-06} \nonumber \\
\end{eqnarray}
\end{small}
where
\begin{eqnarray}
\sin \gamma &=&-\frac{1}{A_{b}}, \ \ \ \ \ \ \ \ \ \ \ \ \cos \gamma =-\frac{1}{A_{a}}.
\end{eqnarray}

\begin{figure*}[htbp]
\includegraphics[clip=true,height=8cm,width=14cm]{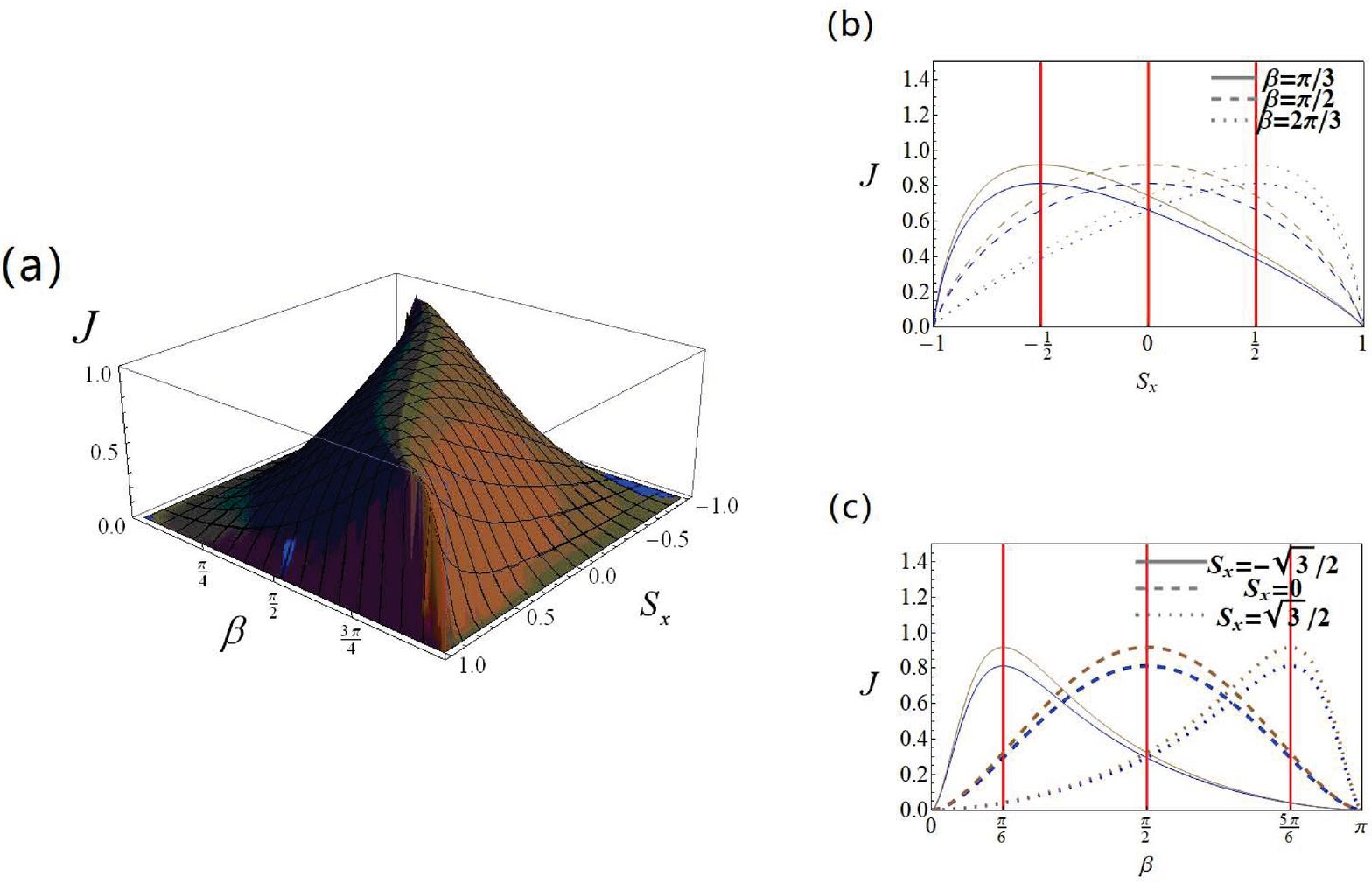}
\caption{(Color online).(a) The CC $\mathcal{J}(\rho^{QD}_{1})$  and
 $\mathcal{J}(\rho^{QD}_{2})$ as functions of the $A$ and $\beta$ with $A=1/3$ are
represented in brown and blue, respectively. (b) The cross section of the 3D surface for $\beta=\pi/3,\pi/2,2\pi/3$, (c) the cross section of the 3D surface for $S_{x}=-\sqrt{3}/2,0,\sqrt{3}/2$. } \label{fig2.eps}
\end{figure*}                                                                                                                                                                                                                    

\begin{figure*}[htbp]
\includegraphics[clip=true,height=8cm,width=14cm]{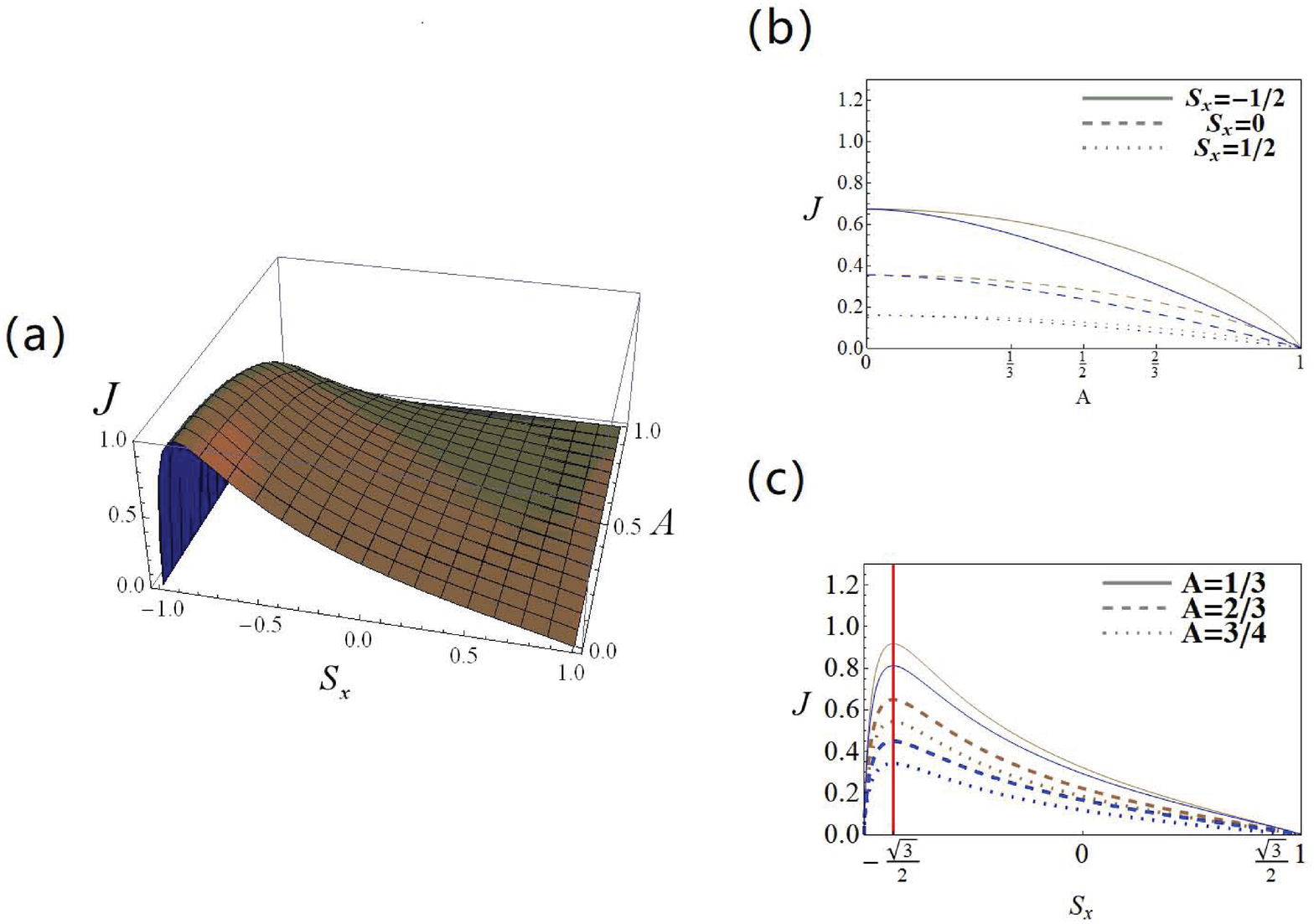}
\caption{(Color online).(a) The CC $\mathcal{J}(\rho^{QD}_{1})$ and
$\mathcal{J}(\rho^{QD}_{2})$ as functions of the $A$ and $S_{x}$ with $\beta=\pi/6$ are
represented in brown and blue, respectively. (b) The cross section of the 3D surface for $S_{x}=-1/2, 0, 1/2$, (c) the cross section of the 3D surface for $A=1/3, 2/3, 3/4$.} \label{fig3.eps}
\end{figure*}                                                                                                                                                                                                                    

\begin{figure*}[htbp]
\includegraphics[clip=true,height=8cm,width=14cm]{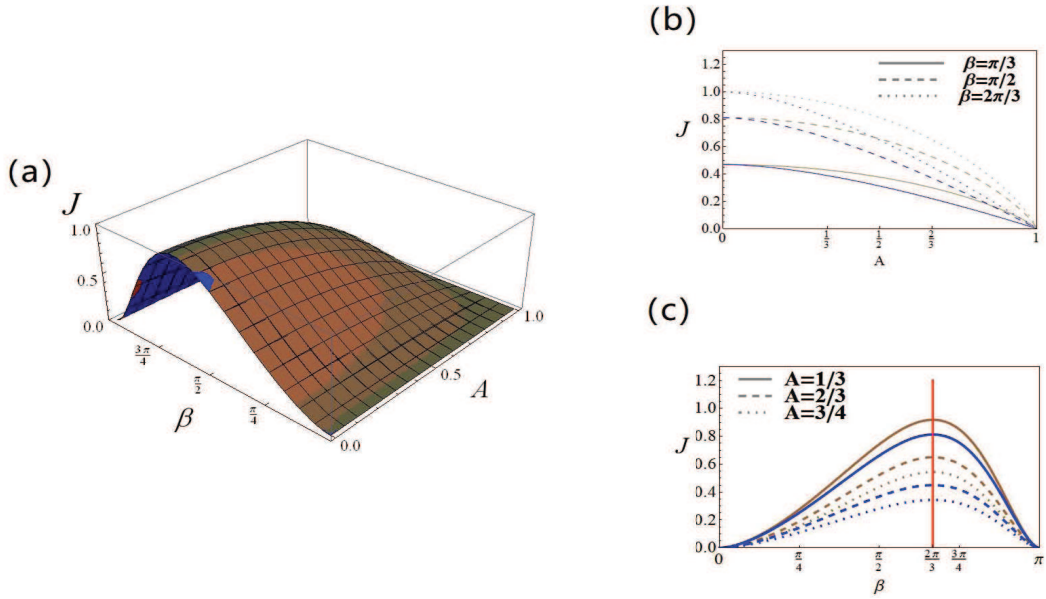}
\caption{(Color online).(a)The CC $\mathcal{J}(\rho^{QD}_{1})$ and
$\mathcal{J}(\rho^{QD}_{2})$ as functions of the $A$ and $\beta$ with $S_{x}=1/2$ are
represented in brown and blue, respectively. (b) The cross section of the 3D surface for $\beta=\pi/3,,\pi/2,2\pi/3$, (c) the cross section of the 3D surface for$A=1/3, 2/3, 3/4$.} \label{fig4.eps}
\end{figure*}                                                                                                                                                                                                                    

Equations (\ref{3eq-04}) and (\ref{3eq-06}) show that the $\mathcal{J}( \rho _{1}^{QD})$ and the $\mathcal{J}( \rho _{2}^{QD})$
are functions of the parameters $S_{x}$, $\beta$ and $A$, respectively. In figure 2, we plot the $\mathcal{J}( \rho _{1}^{QD})$ and the $\mathcal{J}( \rho _{2}^{QD})$ as a function of $S_{x}$ and $\beta$ with the parameter $A = 1/3$ in brown and blue, respectively. From Figure 2, we can obtain the following conclusions.
(1) The values of the $\mathcal{J}( \rho _{1}^{QD})$ and the $\mathcal{J}( \rho _{2}^{QD})$ first increase and then decrease as $S_{x}(\beta)$ increase for a given
$\beta(S_{x})$. (2) The peak appears at $S_{x}=-1/2$ when $\beta=\pi/3$, $S_{x}=0$ when $\beta=\pi/2$, $S_{x}=1/2$ when $\beta=2\pi/3$ in Fig. 2(b); $\beta=\pi/6$ when $S_{x}=-\sqrt{3}/2$, $\beta=\pi/2$ when $S_{x}=0$, $\beta=5\pi/6$ when $S_{x}=\sqrt{3}/2$ in Fig. 2(c). (3) The value of
the CC between the particle and the WPD is zero in the following situations. (a)
The effect of the BS2 for the particle is full transmission(full reflection), corresponding to $\beta= 0(\pi)$. (b) The particle
only travels along the $a(b)$ path, corresponding to
$S_{x} = 1(-1)$. In figure 3(4), we plot the $\mathcal{J}( \rho _{1}^{QD})$ and the $\mathcal{J}( \rho _{2}^{QD})$ as a function of $S_{x}(\beta)$ and $A$ with the parameter
$\beta= \pi/6(S_{x}=1/2)$ in brown and blue, respectively. The blue part shown in Figs. 2(a) and 4(a) indicates $\mathcal{J}( \rho _{1}^{QD})=\mathcal{J}( \rho _{2}^{QD})$. From Figs. 3(b)
and 4(b), one finds that the $\mathcal{J}( \rho _{1}^{QD})$ and the $\mathcal{J}( \rho _{2}^{QD})$ decreases as $A$ increases.
The peak appears at $S_{x}=-\sqrt{3}/2$ when $\beta=\pi/6$ in figure 3(c) and $\beta=2\pi/3$ when $S_{x} = 1/2$ in figure 4(c).
By analyzing Figs. 2$-$4, we can obtain the following conclusions.(1) Under the
same conditions, the relationship of $\mathcal{J}( \rho _{1}^{QD})$ and $\mathcal{J}( \rho _{2}^{QD})$ satisfies
$\mathcal{J}( \rho _{1}^{QD})\geq\mathcal{J}( \rho _{2}^{QD})$. (2)  The maximum of the $\mathcal{J}( \rho _{1}^{QD})$ and $\mathcal{J}( \rho _{2}^{QD})$ can be achieved once $\cos \beta=-S_{x}$. The $\mathcal{J}( \rho _{1}^{QD})$ obtained the maximum
\begin{equation}
\mathcal{J}( \rho _{1}^{QD})_{max} =-\frac{1+A}{2}\log ( \frac{1+A}{2}%
) -\frac{1-A}{2}\log ( \frac{1-A}{2}), \label{3eq-07}
\end{equation}
when $\cos \beta=-S_{x}$.

\begin{center}
\includegraphics{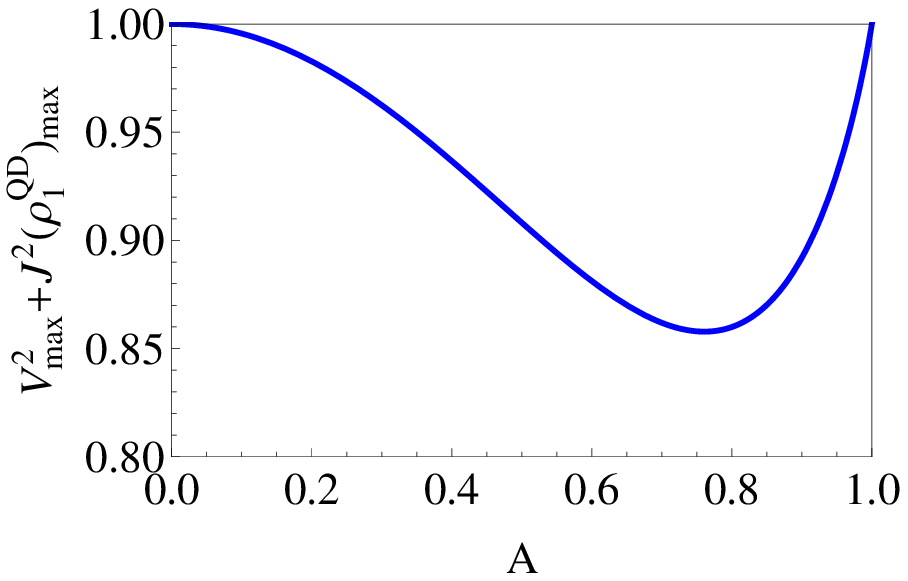}\\[5pt] 
\parbox[c]{15.5cm}{\footnotesize{\bf Fig.~5.} The relationship between the $\mathcal{J}^{2}(\rho _{1}^{QD})_{max}+V^{2}_{max}$ and the A .}
\end{center}

In figure 5, we plot the $\mathcal{J}^{2}( \rho _{1}^{QD})_{max}+V^{2}_{max}$ as a function of A. We find the relationship
\begin{equation}
\mathcal{J}^{2}( \rho _{1}^{QD})_{max}+V^{2}_{max}\leq1 \label{3eq-08}
\end{equation}
for any A. So, we can obtain the complementary relation of the $\mathcal{J}( \rho _{1}^{QD})$ and the $V$
\begin{equation}
\mathcal{J}^{2}( \rho _{1}^{QD})+V^{2}\leq1 \label{3eq-09}.
\end{equation}
Due to the relationship between $\mathcal{J}( \rho _{1}^{QD})$ and $\mathcal{J}( \rho _{2}^{QD})$, we can also obtain the
complementarity relationship between the $\mathcal{J}( \rho _{2}^{QD})$ and the $V$
\begin{equation}
\mathcal{J}^{2}( \rho _{2}^{QD})+V^{2}\leq1 \label{3eq-10}.
\end{equation}
The equals sign hold in equations (\ref{3eq-09}) and (\ref{3eq-10}) in the following situations.(1) The value of
the fringe visibility is one, corresponding to $\cos \beta=-S_{x}$ and $V=1$. This means the state of the WPD satisfies
$|r\rangle=U|r\rangle$, the interference pattern is completely observed.
(2) The value of
the $\mathcal{J}( \rho _{1}^{QD})$ or $\mathcal{J}( \rho _{2}^{QD})$ is one,
corresponding to $\cos \beta=-S_{x}$ and $V=0$. This means the state of the WPD
$|r\rangle$ and $U|r\rangle$ are mutually orthogonal, a perfect
WPI of the particle propagating is achieved.

\subsection {Quantum discord via general MZI}

QD is usually used to denote QC, and QD between the particle and the WPD is defined as the total correlation minus the CC
\begin{equation}
\mathcal{D}(\rho^{QD}) = I(\rho^{QD})-\mathcal{J}(\rho^{QD}) \label{3eq-11},
\end{equation}
here, quantum mutual information $I(\rho^{QD})$ is used to quantify the total correlation
\begin{equation}
I(\rho^{QD})=S(\rho^{Q})+S(\rho^{D})-S(\rho^{QD}) \label{3eq-12},
\end{equation}
where $S$ is the von Neumann entropy, $\rho^{Q}$ and $\rho^{D}$ are the reduced density matrix of $\rho^{QD}$.

(1) Quantum discord for the pure state at the output of MZI

By simple calculations, we can obtain the QD between the particle and the WPD when their final state is in pure state
\begin{equation}
\mathcal{D}( \rho _{1}^{QD})=\mathcal{J}(\rho^{QD}_{1}) \label{3eq-13}.
\end{equation}
So, we obtain the complementary relation
\begin{equation}
\mathcal{D}^{2}( \rho _{1}^{QD})+V^{2}\leq1 \label{3eq-14}.
\end{equation}

(2) Quantum discord for the mixed state at the output of MZI

By the Equation (\ref{3eq-11}), we can obtain that when the final state of the particle and the detector is in mixed state, the
QD
\begin{equation}
\mathcal{D}( \rho _{2}^{QD})=\mathcal{J}( \rho _{1}^{QD})-\mathcal{J}( \rho _{2}^{QD})\label{3eq-15}.
\end{equation}
From equation (\ref{3eq-15}), we find that the total correlation of the particle and the WPD
\begin{equation}
I( \rho _{2}^{QD})=\mathcal{J}( \rho _{1}^{QD}) \label{3eq-16}.
\end{equation}

Liu et al.$[39]$ have demonstrated that $\mathcal{D}(\rho _{2}^{QD})$ and $V$ do not satisfy the complementarity
when $S_{x}= - \cos \beta = 0$ and
the input particle is in pure state. For the more general case we discuss
in this section, when the final state of the particle and the WPD is in mixed state, the $\mathcal{D}(\rho _{2}^{QD})$ and $V$ also do
not satisfy the complementarity.

\section{\label{Sec:4} conclusion}

We have investigated the complementarity of the fringe visibility and the CC or QD in an MZI with one asymmetric BS.
The fringe visibility $V$ exhibits an upper bound $A$
when the particle initially is in pure state and $\cos \beta=-S_{x}$.
We solve the CC $\mathcal{J}(\rho^{QD}_{1})$ and the QD $\mathcal{D}(\rho^{QD}_{1})$
between the particle and the WPD when the final state of them is in pure state,
and find that $\mathcal{D}( \rho _{1}^{QD})=\mathcal{J}(\rho^{QD}_{1})$ and the maximum
of the $\mathcal{J}(\rho^{QD}_{1})$ and the $\mathcal{D}(\rho^{QD}_{1})$ can be achieved when $\cos \beta=-S_{x}$.
The CC $\mathcal{J}( \rho^{QD}_{1})$ or the QD $\mathcal{D}(\rho^{QD}_{1})$ and
the fringe visibility $V$ satisfy the complementarity relationship $\mathcal{J}^{2}(\rho^{QD}_{1})(\mathcal{D}^{2}(\rho _{1}^{QD}))+V^{2}\leq1$.
We also obtain the CC $\mathcal{J}(\rho^{QD}_{2})$ and QD $\mathcal{D}(\rho^{QD}_{2})$
between the particle and the WPD when the
final state of them is in mixed state. $\mathcal{J}(\rho^{QD}_{2})$ can obtain the
maximum once $\cos \beta=-S_{x}$. The complementarity relationship between $\mathcal{J}(\rho^{QD}_{2})$ and $V$ satisfies
$\mathcal{J}^{2}(\rho^{QD}_{2})+V^{2}\leq1$.
It is also found that $\mathcal{J}(\rho^{QD}_{1})(\mathcal{D}(\rho^{QD}_{1}))\geq\mathcal{J}(\rho^{QD}_{2})(\mathcal{D}(\rho^{QD}_{2}))$,
which means that the environment not only induces the optimal measurement basis, but also
reduces the CC and QD between the particles and the WPD.

\end{document}